\documentclass[twocolumn,prb]{revtex4-1}%
\usepackage{amsfonts}
\usepackage{amsmath}
\usepackage{amssymb}
\usepackage{color}
\usepackage{graphicx}

\begin{document}

\title{Bends In Nanotubes Allow Electric Spin Control and Coupling}
\author{Karsten Flensberg$^1$ and Charles M.~Marcus$^{1,2}$}
\affiliation{$^1$Niels Bohr Institute \& Nano-Science Center, University of Copenhagen,
Universitetsparken 5, DK-2100 Copenhagen, Denmark \\ $^2$Department of Physics, Harvard University, Cambridge, Massachusetts 02138, USA}

\begin{abstract}
We investigate combined effects of spin-orbit coupling and magnetic field in carbon nanotubes containing one or more bends along their length. We show how bends can be used to provide electrical control of confined spins, while spins confined in straight segments remain insensitive to electric fields. Device geometries that allow general rotation of single spins are presented and analyzed. In addition, capacitive coupling along bends provides coherent spin-spin interaction, including between otherwise disconnected nanotubes, completing a universal set of one- and two-qubit gates.
\end{abstract}
\maketitle

\section{Introduction}

Electron spins in confined nanostructures show promise as a basis for quantum information processing.\cite{Petta2005,Nowack2007,Hanson2007a,Pioro-Ladriere2008} Among the realizations of spin qubits, gated carbon nanotubes offer a number of attractive features, including large confinement energy and a nearly nuclear-spin-free environment. A novel circumferential spin-orbit coupling in nanotubes, mediated by \textit{s-p} hybridization and inversely proportional in strength to the nanotube radius,  has been investigated experimentally\cite{Kuemmeth2008,Churchill2009b} and theoretically\cite{Ando2000,Huertas-Hernando2009,Chico2009,Jeong2009,Izumida2009} recently. In this paper, we show that circumferential spin-orbit coupling provides a natural means of creating a strong spatial dependence of the magnitude and direction of the effective magnetic field experienced by a spin qubit formed by confining charge in a nanotube. Along bends in the nanotube, the spin qubit couples efficiently to electrostatic gates, allowing spin control and spin-spin coupling, while along straight regions the spin qubit is insensitive to electric fields and is therefore inactive and protected. Related effects of bending modes on spin relaxation have also been considered recently. \cite{Rudner2010}
\begin{figure}[b]
\label{Fig1}
\centerline{\includegraphics[width= 2.8 in]{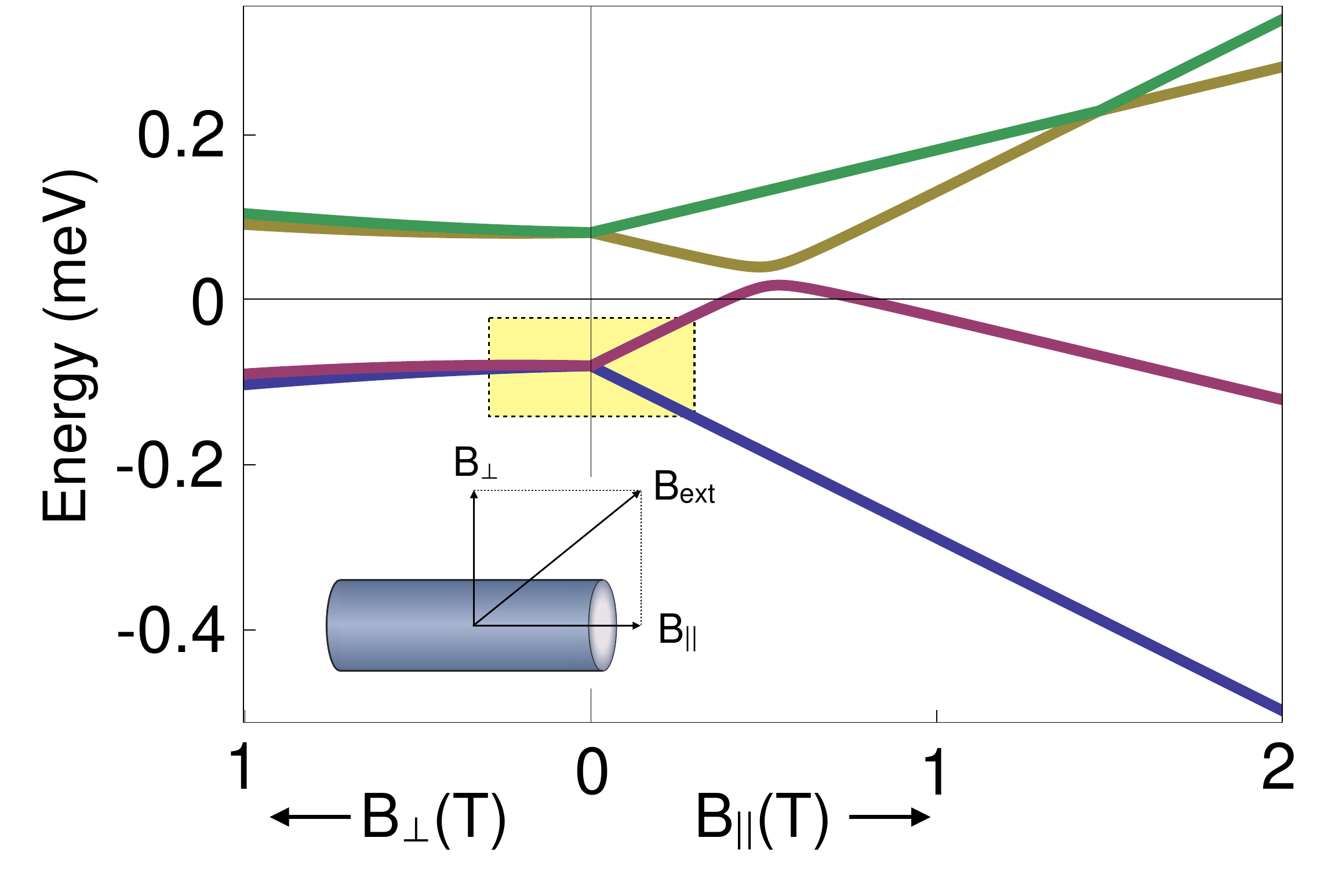}}
\caption{(Color online) Spectrum of low-energy states of a nanotube quantum dot as a function of magnetic field along axial ($B_\parallel$) and transverse ($B_\perp$) directions, for typical device values ($\Delta_{KK'} = 25\,\mu$eV, $\Delta_\mathrm{SO} = 170\,\mu$eV) from Ref.~\onlinecite{Churchill2009b}.  The (yellow) box marks the low-field Kramers doublet, or qubit, for single-electron (mod four) occupancy. Spin-orbit coupling leads to a large effective g factor for axial fields, $B_\parallel$, and small effective g factor for transverse fields, $B_\perp$. Inset shows axial and transverse projections of applied field.}
\end{figure}

The spin, or quantum two-level system, that forms the physical qubit is a Kramers doublet in a nanotube quantum dot containing an odd number of electrons, in the low-magnetic-field regime. As illustrated in Fig.~1, splitting of these doublets in a magnetic field depends on the direction of the field with respect to the nanotube axis. This is the key observation of our analysis: in tubes with bends, the angle between the tube axis and the applied magnetic field depends on position along the tube. This dependence couples position and spin, allowing electric fields to control spin and create spin-spin coupling. In straight segments, changes in position do not change the angle between the field and the nanotube axis, and so this coupling vanishes. For use as a qubit, relaxation of the low-field Kramers doublet is suppressed due to time-reversal symmetry, in contrast to the qubit formed at the high-field crossing\cite{Bulaev2008,Rudner2010} (at 1.4 T in Fig.~1), consistent with experiment.\cite{Churchill2009b}

\section{The Kramers qubit}

We start by analyzing the spectrum of a quantum dot confined along a bend in a nanotube. The geometry of the system can be described in terms of local (primed) coordinates, $x'$, perpendicular to the nanotube axis, and $y'$ along the nanotube axis (see Fig.~2) at the position of the dot. For bend radius $r$ much greater than the interatomic distance, the nanotube band structure is described by that of a locally straight tube,\cite{Ando2005} including spin-orbit interaction.\cite{Ando2000,Huertas-Hernando2009,Chico2009,Jeong2009,Izumida2009}
For a nanotube quantum dot\cite{Bulaev2008} of length $L\ll r$, the effective Hamiltonian to leading order in $L/r$ is
\begin{eqnarray}
H&=&-\frac{1}{2} \left(\tau_3 \Delta_\mathrm{SO}\boldsymbol{\sigma}\cdot\hat{\mathbf{y}}'+\tau_1\Delta _{KK^{\prime }}\right)\notag
\\&&+g_s\mu_B \boldsymbol{\sigma}\cdot\mathbf{B}_\mathrm{ex}
+\tau_3g_\mathrm{orb}\mu_B \mathbf{B}_\mathrm{ex}\cdot\hat{\mathbf{y}}',
\label{eq:Hmethods}
\end{eqnarray}
where $\sigma _{i}$ and  $\tau_{i}$ are Pauli matrices in spin and valley space, respectively, $\Delta_\mathrm{S0}$ is spin-orbit coupling energy, and $\Delta _{KK^{\prime }}$ is a valley mixing term due to substrate, contacts, gates, or any disorder that breaks the crystal symmetry. The first two terms describe the two Kramers doublets, while the last term describes the coupling to magnetic fields of spin and orbital moments. Note that orbital moments are always along the nanotube axis unit vector $\hat{\mathbf{y}}'$. We consider only planar devices with magnetic fields applied in plane of the bent nanotube, but this restriction can be readily generalized to bends and fields out of the plane.

The magnetic field dependence of a nanotube quantum dot, including effects of spin-orbit coupling, is shown in Fig.~1 for typical parameters for small-gap semiconducting nanotubes. Two parameters characterizing the nanotube device are the spin-orbit energy scale, $\Delta_{\rm SO}$, and $\Delta_{KK'}$, which characterizes the mixing of $K$ and $K'$ valleys due to disorder on length scales comparable to or smaller than the nanotube radius, $R$. Fourfold degeneracy at zero applied field, $B_{\rm ext} \equiv|{\mathbf B}_{\rm ext}|= 0$, is lifted by $\sqrt{\Delta_{\rm SO}^2 + \Delta_{KK'}^2}$, typically $\sim 0.4$\,meV, giving two doublets that are well separated at low temperatures. Either doublet may constitute a qubit in the present scheme, depending on occupancy of the dot. Here, we concentrate on the lower pair, appropriate for a single confined electron, in the low-field regime (boxed region in Fig.~1), away from the anticrossing of different orbital states.

Diagonalizing the above 4$\times 4$ Hamiltonian and projecting onto the lowest two eigenstates yields an effective spin-1/2 system, which is our qubit. It has an anisotropic g factor described by the Hamiltonian
\begin{equation}
H^{\ast } =\frac{1}{2}\mu _{B}^{{}}~\mathbf{s}^{\ast }\mathbf{\cdot g}
\cdot \mathbf{B}_{\mathrm{ext}} \label{tensor},
\end{equation}
where $\mathbf{s}^{\ast }$ are Pauli matrices, $\mu_B$ is the Bohr magneton, and $\mathbf{g}$ is the gyromagnetic tensor. In terms of local nanotube coordinates,
\begin{equation}\label{HKramerscoor}
H^{\ast } =\frac{1}{2}\mu _{B}\left( g_{\perp }s_{\perp }^{\ast }B_{\perp
}+g_{\parallel }s_{\parallel }^{\ast }B_{\parallel }\right),
\end{equation}
where $\perp$ and $\parallel$ refer to components of the vectors in Eq.~\eqref{HKramerscoor} along $x'$ and $y'$, respectively. Components of $\mathbf g $ can be expressed in terms of nanotube parameters,
\begin{eqnarray}
g_{\parallel } &=&g_{s}+\frac{2g_{\mathrm{orb}}\Delta _{\rm SO}}{\sqrt{\Delta_{KK^{\prime }}^{2}+\Delta _{\rm SO}^{2}}},  \label{gpar}\\
g_{\perp } &=&\frac{g_{s}\Delta _{KK^{\prime }}}{\sqrt{\Delta _{KK^{\prime
}}^{2}+\Delta _{\rm SO}^{2}}}\label{gperp},
\end{eqnarray}%
where $g_s\sim2$ is the spin g factor and $g_{\rm orb}$ is the orbital g factor, with $g_{\rm orb}/g_s \sim 10$ for typical nanotubes.  We emphasize that because the coordinates are local, changes in confinement position along a bend will change the directions and magnitudes of field components $B_\parallel$ and $B_\perp$ for fixed external field. Eqs.~\eqref{gpar} and \eqref{gperp} show how spatial inhomogeneity in $\Delta_{KK'}$ can also couple spin to position. When this inhomogeneity is small compared to either $\Delta_{KK'}$ or $\Delta_\mathrm{SO}$ this effect is weak.

We introduce the effective field, $\mathbf{B}^{\ast }=\mathbf{g}\cdot \mathbf{B}_{\mathrm{ext}}/g_{s}$, felt by an electron spin, including spin-orbit effects, as a function of position along the nanotube. Variation in the magnitude and direction of $B^*$ along a bend are shown in Fig.~2 for realistic device parameters.
\begin{figure}[t]
\centerline{\includegraphics[width=3.2 in]{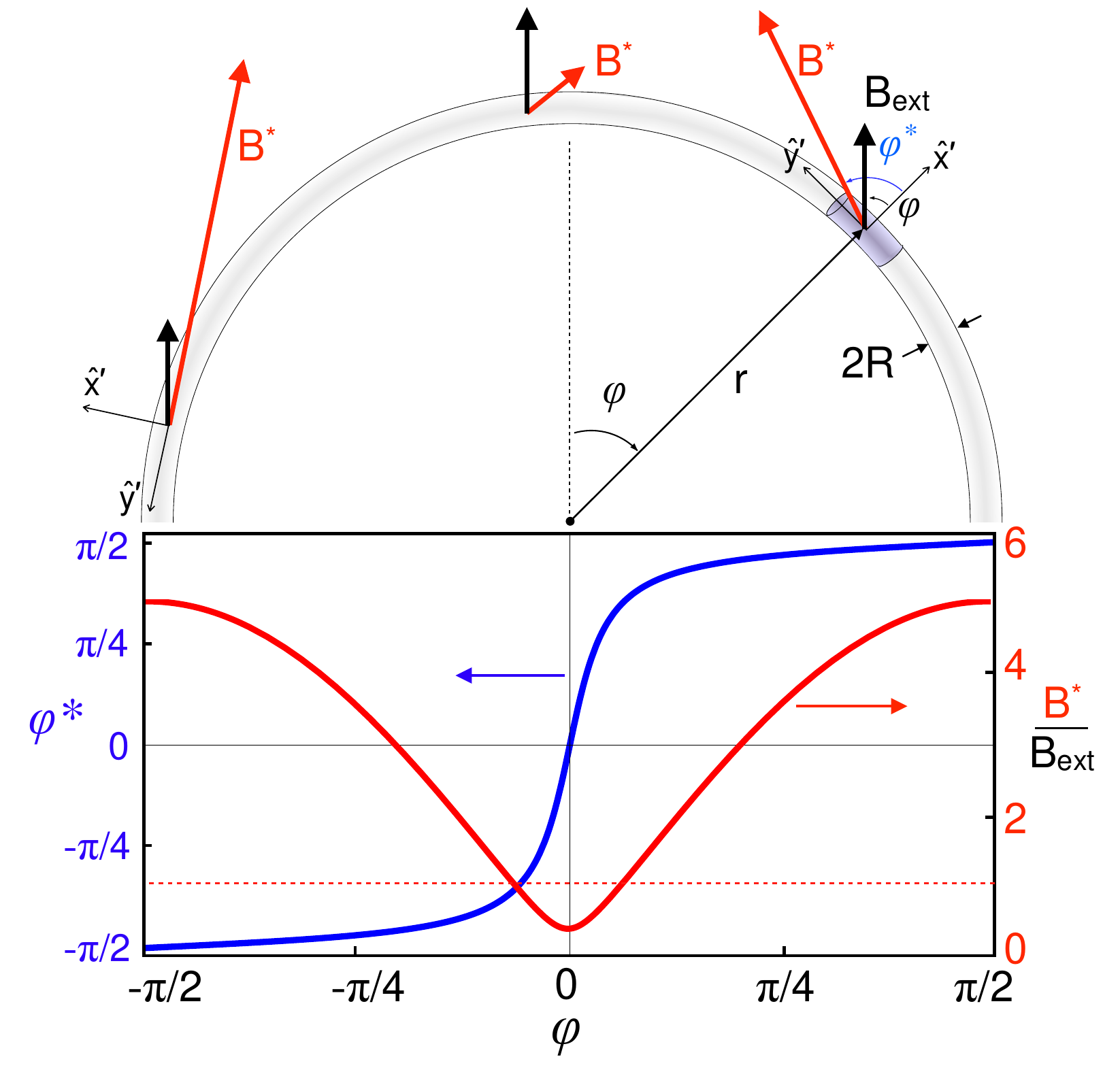}}\caption{(Color online) {Effective fields along a bend.} (a) Illustration of how the magnitude, $B^*$, (red) and angle, $\varphi^*=\angle (\mathbf{B}^*,\mathbf{x}')$, (blue) relative to local $(x',y')$ coordinates of the local effective magnetic field depend on angle, $\varphi$, between the applied field, $\mathbf{B}_{\rm ext}$ (black) and the local transverse direction. Nanotube radius, $R$, and bend radius, $r$, are indicated. (b) Values for angle (blue) and normalized magnitude (red) of effective field, $\mathbf{B}^*$ as a function of angle, $\varphi$, for realistic parameters,\cite{Churchill2009b} with $g_\parallel = 10$ and $g_\perp = 1$. Dashed line (red) indicates $B^*/B_{\rm ext}$ = 1.}
\end{figure}
Because the $\mathbf{g}$ tensor in local coordinates, $\mathbf{g}'$, is diagonal, the effective field is found by $\mathbf{B}^{\ast }=\mathbf{R}_{\varphi }^{-1}\cdot \mathbf{g}^{\prime }\mathbf{\cdot R}_{\varphi }^{{}}\mathbf{\cdot B}_{\mathrm{ext}}^{{}}/g_{s}$,
where $\mathbf{R}_{\varphi }^{{}}$ is the matrix that rotates $\mathbf{B}_{\mathrm{ext}}$ to the local nanotube coordinates. The effective field is
\begin{eqnarray}
\mathbf{B}^{\ast } &=&\frac{g_{\parallel }\!+\!g_{\perp }\!-\!\left(
g_{\parallel }\!-\!g_{\perp }\right) \cos 2\varphi }{2g_{s}}\mathbf{B}_{\mathrm{ext}}^{{}}  \notag \\
&&+\mathbf{\hat{z}\times B}_{\mathrm{ext}}^{{}}\frac{g_{\parallel}\!-\!g_{\perp }}{2g_{s}^{{}}}\sin 2\varphi .
\end{eqnarray}
This formula forms the basis for the discussion in Figs.~2 and 4.
\begin{figure}[t]
\centerline{\includegraphics[width=2.8 in]{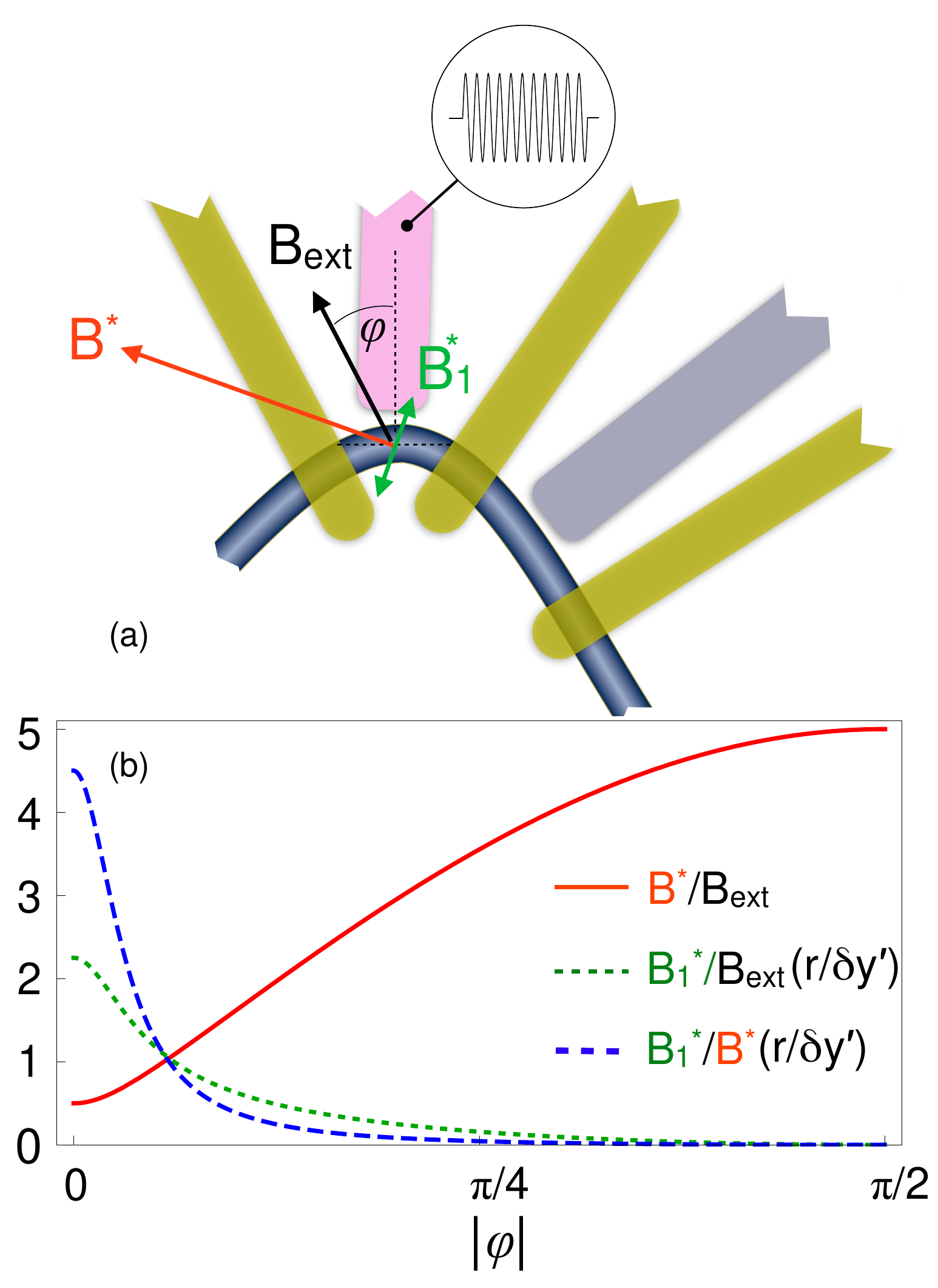}}\caption{(Color online) {Gate-driven spin resonance on bend.} (a) The
geometry for electric dipole spin resonance (EDSR) for quantum dot confined along a bend. Spin in the quantum dot in the adjacent straight region is not sensitive to electric fields. An oscillating gate voltage produces an oscillating transverse field, $B_1^*$, which on resonance drives efficient qubit rotation around $B^*$.  (b) Normalized static field, $B^*$, (red) and oscillating transverse field, $B_{1}^*$ as a function of angle, $\varphi$, of the applied field measured in local dot coordinates, for $g_{\parallel}=10$ and $g_{\perp}=1$. Resonance frequency, $g_s \mu_B B^*/h$, depends on dot position, allowing frequency tuning and multiplexing. The scale of the transverse field scales with $\delta y'/r$, the ratio of the axial excursion due to the oscillating gate voltage to bend radius, $r$.}
\end{figure}

\section{Electric dipole spin resonance}

The sizable variation of $\mathbf{B}^*$ along the bend suggests several applications that involve both finite and infinitesimal gate-induced motion.  As a first example, we consider electric dipole spin resonance (EDSR) using an oscillating gate voltage, as illustrated in Fig.~3. We consider a single electron confined by gates to a region containing a bend. A second electron confined in an adjacent dot may be used to detect spin rotation via Pauli blockade.\cite{Hanson2007a} An external field is applied at an angle $\varphi$ to the local transverse ($x'$) direction, and a gate voltage, oscillating at frequency $f$, causes the center position of the dot to move by an amount $\delta y'$. This motion modifies the orbital coupling to the applied field, while Zeeman coupling is unchanged. Small displacements $\delta y'$ of the dot position then result in a perturbation $\delta H=(\delta y'/r)(\partial H/\partial\varphi)$ of the Hamiltonian \eqref{eq:Hmethods}. Only the tangent vector $\hat{\mathbf{y}}'$ in \eqref{eq:Hmethods} depends on $\varphi$, with derivative
$\partial \hat{\mathbf{y}}'/\partial \varphi=\hat{\mathbf{x}}'$. Again projecting onto the lowest Kramers doublet, the modulation of the effective field becomes $\delta \mathbf{B}^*=(\delta y'/r)\mathbf{B}^{**}$, where
\begin{equation}\label{dBstarstar}
    \mathbf{B}^{**}=\delta\mathbf{g}\cdot\mathbf{B}_\mathrm{ext}/g_s,
\end{equation}
and
\begin{equation}
    \delta\mathbf{g}=
    \frac12 \left(
      \begin{array}{cc}
        0 & 0 \\
        g_\parallel-g_s & 0 \\
      \end{array}
    \right).
\end{equation}
Therefore the oscillatory part of the effective field points along the nanotube axis, $\hat{\mathbf{y}}'$, which in general is not aligned with $\mathbf{B}^*$, but has components both along $\mathbf{B}^*$, with magnitude $\delta B^*$, and transverse to $\mathbf{B}^*$, with magnitude $B_1^*$, given by
\begin{align}
\frac{\delta B^{\ast }}{B^*} &=\left(\frac{\delta y^{\prime }}{r}\right)
\frac{g_\parallel\left(g_{\parallel }-g_{\perp }\right) \sin(2\varphi)}{2b^{2}(\varphi )},\label{dBstar}
\\
\frac{B_{1}^{\ast }}{B^{\ast }}&=\left(\frac{\delta y^{\prime }}{r}\right)
\frac{g_\perp\left( g_{\parallel}-g_{\perp }\right) \cos^2(\varphi)}{2b^{2}(\varphi )},\label{B1star}
\end{align}
where $b(\varphi )=\sqrt{g_{\perp }^{2}\cos ^{2}\varphi +g_{\parallel }^{2}\sin
^{2}\varphi}=g_sB^*/B_\mathrm{ext}$. For a reasonable nanotube bend radius, $r=1\,\mu$m, and gate-induced dot motion, $\delta y' \sim 1$\, nm, the plots in Fig.~3 indicate a transverse field of order $B_1^*\sim 10^{-4}$\,T for $B_{\rm ext} = 100$\,mT, an applied field for which the qubit remains well defined (Fig.~1). At the resonance frequency, $f = \frac{1}{2} g_s \mu_B B^*/h$ ($h$ is Planck's constant), the electron spin will precess at the Rabi frequency, $g_s\mu_B B_1^*/h$, which exceeds several MHz.

Note from Eq.~\eqref{B1star} that the transverse oscillating field, $B^*_1$, vanishes in the absence of valley mixing. For weak valley mixing, $\Delta_{KK'}\ll\Delta_\mathrm{SO}$, the maximal ratio $B_1^*/B^*\propto g_\parallel/g_\perp$ is obtained at $\varphi=0$, i.e., when the applied field is transverse to tube.

Cross coupling of ac gate voltages to dots in nearby straight regions of the nanotube (as in the example in Fig.~3) will not effect spins there. Moreover, adjacent dots also in bent regions (with the same or different $r$) will have different resonant frequencies---$f$ depends on position along a bend---and so will be relatively insensitive to the oscillating gate.

This example illustrates how modest bends and ac gate voltages are capable of generating efficient and selective spin rotation with transverse field strengths comparable to existing few-spin EDSR schemes.\cite{Golovach2006,Nowack2007,Laird2007,Pioro-Ladriere2008}

\section{Fast spin rotation via non-adiabatic passage through bends}

As  a second example of spin manipulation, we consider the geometry in Fig.~4(a), consisting of two straight segments on either side of a single bend, with radius of curvature $r$, forming an angle $2\theta$. Two quantum dots, denoted $a$ and $b$, are defined by gates on the straight segments of this ``coat hanger" shape, and the external field is applied in the plane, at an angle $\varphi$ with respect to the symmetry axis. Effective fields ${\mathbf B} ^*_a$ and ${\mathbf B}^*_b$ in the two dots differ in both magnitude and direction. In particular, the angle $\eta$ between ${\mathbf B}^*_a$ and ${\mathbf B}^*_b$, given by
\begin{equation}
\sin \eta =\frac{\left[ (g_{\parallel }-g_{\perp })^{2}\cos (2\theta
)+(g_{\perp }^{2}-g_{\parallel }^{2})\cos (2\varphi )\right] \sin (2\theta )%
}{2b(\theta +\varphi )b(\theta -\varphi )},
\end{equation}
can reach $\eta = \pi/2$ for realistic device parameters. Fig.~4(b) shows $\eta$ as a function of bend angle, $\theta$, for two values of g-factor anisotropy, $g_\parallel/g_\perp$, one that does and one that does not exceed the critical value, $g_\parallel/g_\perp = 5.87...$, above which the condition ${\mathbf B}^*_a \perp {\mathbf B}^*_b$ can be met for two values of $\theta$. In particular, typical nanotubes, with $g_\parallel/g_\perp \sim 10$, easily allow this orthogonality condition.

The coat hanger geometry provides nonresonant qubit rotation when an electron is moved nonadiabatically from one dot to the other. Because the precession field is the same order as the quantizing field, precession rates, $\sim g_s \mu_B B^*_a/h$, are typically two to three orders of magnitude faster than the EDSR device described above, allowing nanosecond $\pi$ rotations. As an example, for ${\mathbf B}^*_a \perp {\mathbf B}^*_b$ and $\varphi < \theta$, a qubit initialized in dot $a$ and moved nonadiabatically to dot $b$ will rapidly precess around ${\mathbf B}^*_b$, at a frequency $g_s \mu_B B^*_b/h$. At some point along the passage, ${\mathbf B}_{\rm ext}$ will be purely transverse to the nanotube axis. At this point, the condition for nonadiabatic passage becomes very liberal, only requiring a passage rate faster than $g_\perp\mu_B B_{\rm ext}/h$. From Eq.~\eqref{gperp}, $g_\perp<g_s$, which makes the minimum gap small. Experimental values $\Delta_{KK'} = 25\,\mu$eV and $\Delta_{\rm SO} = 170\,\mu$eV give $g_\perp = 0.15$\,$g_s$. Using $B_{\rm ext}\sim 10$\,mT allows pulse transition occurring in under 10 ns to be considered nonadiabatic, a criterion that is readily achieved with standard arbitrary waveform generators and coaxial cryogenic wiring.

A notable feature of the coat hanger geometry is that the electron spends nearly all of its time---including during rotation---in straight regions of the tube, where stray electric fields do not cause inadvertent qubit rotation; only during the brief non-adiabatic passage from one straight region to another is the qubit on a bend and therefore sensitive to decoherence due to electrical noise. A single bend (as in Fig. 4a) allows spin rotation around a single axis. A nanotube with two bends (for instance, in the shape of the letter N) allows rotation around two axes, and thus arbitrary qubit rotation.

\begin{figure}[t]
\centerline{\includegraphics[width=3 in]{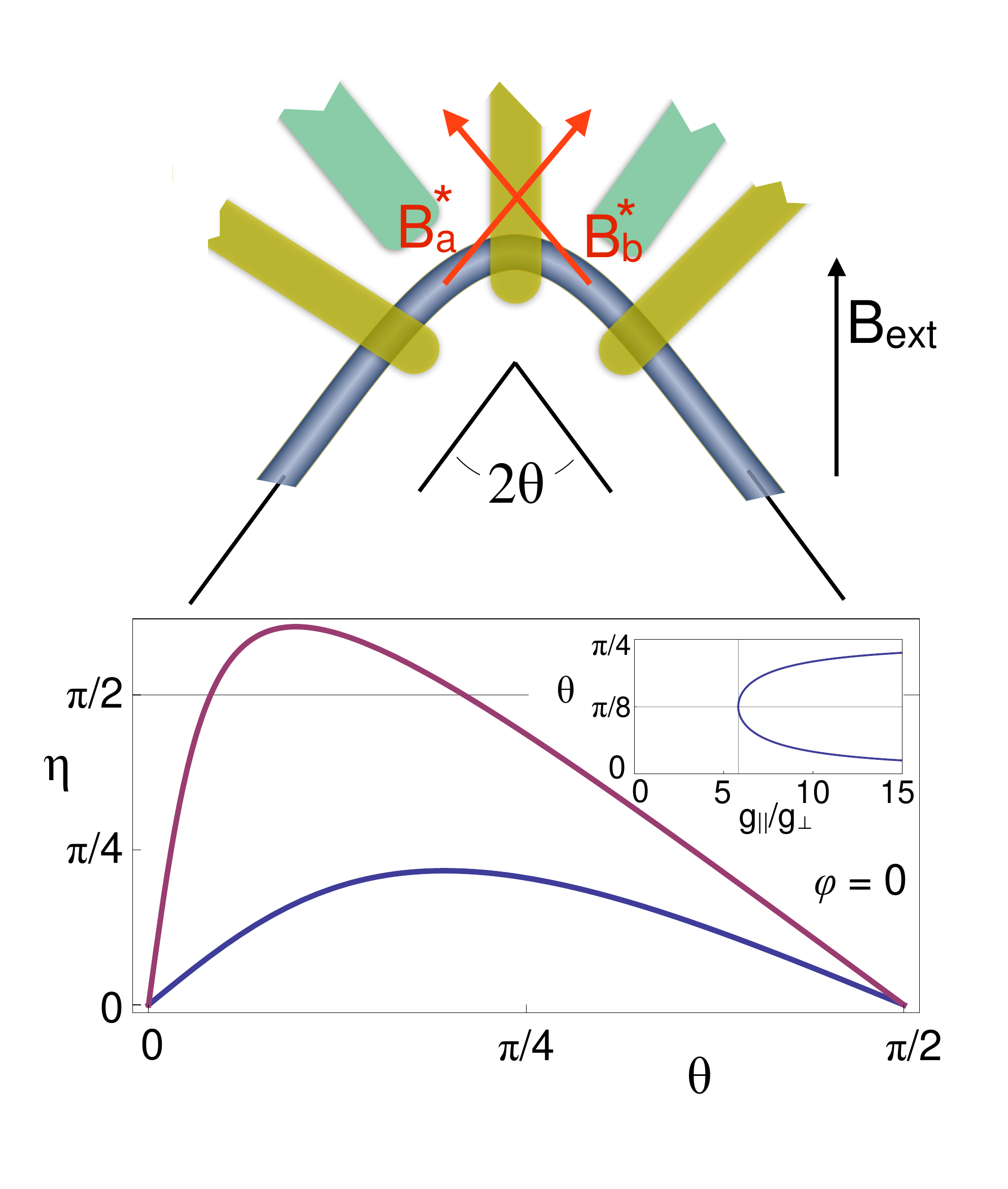}}\caption{(Color online) {Moving between straight segments provides rapid spin rotation.} (a) Two quantum dots (labeled $a$ and $b$) on straight segments with a bend of angle $2\theta$ between them. This geometry provides fast qubit rotation without an oscillating field, while keeping spins in straight segments protected from stray electric fields. A spin quantized in the effective field $\mathbf{B}_a^*$ of dot $a$, when moved nonadiabatically to dot $b$, encounters an effective field, $\mathbf{B}_b^*$, that makes a sizable angle, $\eta$, with $\mathbf{B}_a^*$. The spin then precesses about $\mathbf{B}_a^*$ at a frequency $g_s \mu_B B_b^*/h$ before being returned nonadiabatically to dot $a$.  Shown is the symmetric case, $\varphi = 0$, where the applied field is transverse to the nanotube axis at its apex. The condition for nonadiabatic transfer is easily met for small $g_\perp$ (see text).  (b) The angle, $\eta$, between effective field directions in dots $a$ and $b$ as a function of the bend angle, $\theta$, for $\varphi = 0$ and $g_\parallel/g_\perp=2$ (lower,blue) and $g_\parallel/g_\perp= 10$ (upper,purple). Inset: for $g_\parallel/g_\perp> 5.8$ pairs of values of $\theta$ for which effective fields in $a$ and $b$ are orthogonal, $\eta=\pi/2$.} %
\end{figure}

\section{Spin-spin interaction via capacitative coupling}

The coupling of spin and position also provides a natural mechanism for spin-spin interaction using a capacitive gates or resonant cavities.\cite{Flindt2006,Trif2007} This allows nonlocal two-qubit interaction in a single nanotube by coupling adjacent or non-adjacent qubits using gates between multiple bends, as well as providing two-qubit interaction between different nanotubes. Coherent coupling of spins in different nanotubes using gated bends solves an important challenge of nanotube-based quantum information systems of how to move quantum information through networks or arrays of multiple tubes.
\begin{figure}[b]
\centerline{\includegraphics[width=2.5 in]{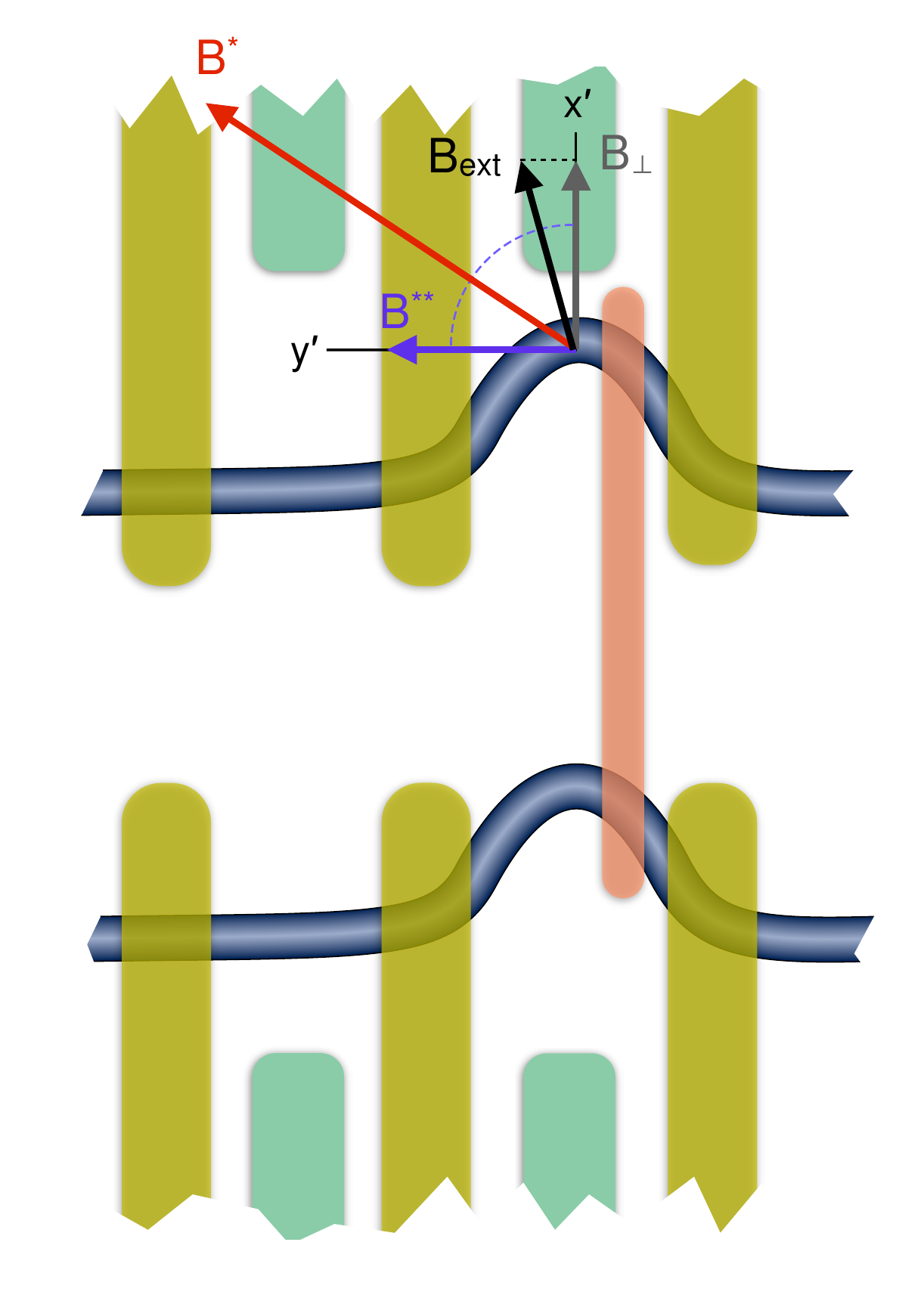}}\caption{(Color online) Schematic of electrostatic coupling via a common gate (red) between two quantum dots formed along bends in two nanotubes. Spin-spin interaction in the quantum dots creates a two-qubit gate that is part of a universal gate set. Applied field, $\mathbf{B}_{\rm ext}$, and local effective fields, $\mathbf{B}^*$ and $\mathbf{B}^{**}$, along with local axes, $x'$ and $y'$, are  shown for one of the two nanotubes.}
\end{figure}

Previous work has demonstrated capacitive coupling between separated quantum dots on the same nanotube,\cite{Churchill2009b} as well as between two quantum dots on different Si/Ge nanowires,\cite{Hu2007} and between a nanotube quantum dot and a single-electron transistor,\cite{Biercuk2006} all over distance scales of $\sim 1\,\mu$m. The coupling between dots is mediated by a relatively large metallic gate whose self-capacitance $C_g$ typically exceeds the self-capacitances $C_1$ and $C_2$ of the dots. In this configuration (see Fig.~5), the electrostatic interaction energy, $U_{12}$, is approximately given by $U_{12} \approx e^2(C_{1g}C_{2g})/(C_{1}C_gC_{2})$ where $C_{1g}$ and $C_{2g}$ are the mutual capacitances of the dots to the coupling gate. Typical values from previous experiments give $U_{12} \sim 0.01\, U$, where $U\sim e^2/C_1\sim e^2/C_2$ is the Coulomb charging energy of the individual nanotube dots, typically $\sim\,$5~meV.

The two-qubit Hamiltonian is $H_{1}+H_{2}+H_{12}$, where $H_1$ and $H_2$ are the single dot Hamiltonians as in Eq.~(1), and where the interaction term, $H_{12}=U_{12}(y_{1}^{\prime },y_{2}^{\prime })$, depends on the position of the electrons within the two dots. For small displacements, the linear terms of $H_{1}+H_{2}$ are
\begin{equation}\label{Hm}
H_1^{(1)}+H_2^{(1)}=\frac{\delta y_{1}^{\prime }}{r_{1}}\left( \mathbf{s}_{1}^{\ast
}\cdot \mathbf{B}^{**}_1\right) +\frac{\delta y_{2}^{\prime }}{r_{2}}\left(\mathbf{s}_{2}^{\ast }\cdot \mathbf{B}^{**}_2\right) ,
\end{equation}
where $\mathbf{B}^{**}_{i}$ is the field defined in \eqref{dBstarstar} for dot $i$. To second order in the dot displacements, the spin interaction becomes
\begin{equation}\label{H12star}
H_{12}^{\ast }=K_{12}\left( \mathbf{s}_{1}^{\ast }\cdot \mathbf{B}^{**}_1\right) \left( \mathbf{s}_{2}^{\ast }\cdot \mathbf{B}^{**}_2\right) ,
\end{equation}
where
\begin{equation}
K_{12}\!=\!\frac{2\mu _{B}^{2}}{r_{1}r_{2}}\,\,\mathrm{Im}\int_{-\infty }^{0}\frac{dt}{\hbar }e^{t0^{+}}\left\langle
\delta y_{1}^{\prime }(t)\delta y_{2}^{\prime }\right\rangle _{B=0}.
\end{equation}
Inserting Eq.~\eqref{dBstarstar} into Eq.~\eqref{H12star} leads to
\begin{equation}\label{eq:H12star}
H_{12}^{\ast }=K_{12}^{\ast }\left(s_{\parallel ,1}^{\ast }B_{\perp ,1}\right) \left(s_{\parallel ,2}^{\ast }B_{\perp ,2}\right),
\end{equation}
with
\begin{equation}
K_{12}^{\ast }=\left( \frac{g_{\parallel }-g_{\perp }}{2g_{s}}\right)^{2}
\frac{\hbar ^{4}\mu _{B}^{2}}{4m^{2}(\Delta E)^{4}r_{1}r_{2}}\frac{\partial
^{2}U_{12}}{\partial \left( \delta y_{1}^{\prime }\right) \partial (\delta
y_{2}^{\prime })},
\end{equation}
for identical parabolic confinement potentials. It has the expected dependences on g-factor anisotropy and radii of curvature of the two bends, $r_1$ and $r_2$. It also depends on the sensitivity of $U_{12}$ to differential motion along the nanotubes, $\delta y'_1$ and $\delta y'_2$. In this expression $m$ is the effective electron mass and $\Delta E$ is the characteristic level spacing in the two quantum dots, which together characterize the stiffness of the confining potential to spin-dependent forces.

This form, with parallel spin component coupled to transverse applied field components is a consequence of locally circular motion along the bends, where changes in field components (which quantize the spin direction) upon infinitesimal motion are transverse to the field components themselves. Expressing  $H^*_{12}$ in terms of the fields $\mathbf{B}^{**} = B_\perp\hat{\mathbf{y}}'$ (Fig.~5) yields a transverse-Ising-like form, which is known to generate spin entanglement between the coupled dots. Two applications of such a gate in combination with single qubit rotations generates a CNOT gate and therefore, together with general single-qubit rotations, constitutes a universal gate set.\cite{Barenco1995}

Realistic values for $K_{12}^\ast$ can be estimated by noting that the dependence of $U_{12}$ on $\delta y'_1$ and $\delta y'_2$, reflects the dependence of mutual capacitances $C_{1g}$ and $C_{2g}$ on dot motion. The characteristic scale of this geometrical dependence is the dot length, $L$, giving the estimate
\begin{equation}
\frac{\partial ^{2}U_{12}}{\partial \left( \delta y_{1}^{\prime }\right)
\partial (\delta y_{2}^{\prime })}\sim \frac{\partial C_{1g}}{\partial
\left( \delta y_{1}^{\prime }\right) }\frac{\partial C_{2g}}{\partial \left(
\delta y_{2}^{\prime }\right) }\frac{1}{C_{1}C_{2}C_{g}} \sim \frac{U_{12}
}{L^{2}}.
\end{equation}
The stiffness, characterizing changes in dot position in response to spin-dependent electrostatic forces, can similarly be estimated by replacing the oscillator length $\ell =(\hbar ^{2}/m\Delta E)^{1/2}$ with the dot length $L$, giving
\begin{equation}
K^*_{12}\sim U_{12}\left( \frac{g_{\parallel }-g_{\perp }}{g_{s}}\right)
^{2}\left( \frac{\mu _{B}}{\Delta E}\right) ^{2}\frac{L^{2}}{4r_{1}r_{2}}.
\end{equation}
Using representative experimental values for coupling strength $U_{12}\sim0.01\,U\sim100\,\mu$eV, g-factor anisotropy $(g_{\parallel}-g_{\perp})/g_{s}\sim10$, level spacing $\Delta E\sim 5$\,meV, and dot size $L\sim0.3\,\mu$m, and taking reasonable values for bend radii $r_1\sim r_2 \sim 0.3\,\mu$m, yields the estimate $K_{12}^*\sim 0.1 \,\mu$eV/T$^2$. For applied fields of 100 \,mT, this strength of coupling allows two-qubit operations on time scales of $\sim1\,\mu$s, which is considerably faster than the anticipated coherence time (which, however, has not yet been measured). Gate operation time can likely be reduced further by decreasing $C_g$, bend radii, or level spacing.

\section{Conclusions}

In summary, the combination of spin-orbit coupling and curved geometry\cite{Belov2005} allows qubit novel control schemes using electric gate manipulation. Notwithstanding the ability to control spin using electric fields in nanotubes with bends, spins confined to straight regions of the nanotube electrons are immune to electrical noise. Bends also allow spin-spin interaction between capacitively coupled nanotubes, providing an entangling transverse-Ising-like two-qubit gate, which along with full single-qubit rotations, provided by nanotubes with two bends, constitutes a universal set of gates.

Various methods for creating nanotubes with bends have been demonstrated. These include growth techniques that yield serpentine nanotubes with multiple bends\cite{Geblinger2008} and manipulation, for instance using an atomic-force microscope,\cite{Postma2001,Bozovic2001,Biercuk2004} following growth.

\acknowledgements
We thank H. Churchill, P. Herring, F. Kuemmeth, D. Loss, J. Paaske, E. Rashba, A. S\o rensen and J. Taylor for discussions. This work was supported in part by the National Science Foundation under grant no.~NIRT 0210736, NRI-INDEX, US Department of Defense, and The Danish Council for Independent Research $\vert$ Natural Science (FNU).

%\bibliography{spin}

\begin{thebibliography}{10}

\bibitem{Petta2005}
J.~R. Petta {\it et~al.}, Science {\bf 309},  2180  (2005).

\bibitem{Nowack2007}
K.~C. Nowack, F.~H.~L. Koppens, Y.~V. Nazarov, and L.~M.~K. Vandersypen,
  Science {\bf 318},  1430  (2007).

\bibitem{Hanson2007a}
R. Hanson, J.~R. Petta, S. Tarucha, and L.~M.~K. Vandersypen, Rev.~Mod.~Phys.
  {\bf 79},  1217  (2007).

\bibitem{Pioro-Ladriere2008}
M. Pioro-Ladri\`{e}re {\it et~al.}, Nature Physics {\bf 4},  776  (2008).

\bibitem{Kuemmeth2008}
F. Kuemmeth, S. Ilani, D.~C. Ralph, and P.~L. McEuen, Nature {\bf 452},  448  (2008).

\bibitem{Churchill2009b}
H.~O.~H. Churchill {\it et~al.}, Phys.~Rev.~Lett. {\bf 102},  166802  (2009).

\bibitem{Ando2000}
T. Ando, J. Phys. Soc. Jpn. {\bf 69},  1757  (2000).

\bibitem{Huertas-Hernando2009}
D. Huertas-Hernando, F. Guinea, and A. Brataas, Phys.~Rev.~Lett. {\bf 103},  146801 (2009).

\bibitem{Chico2009}
L. Chico, M.~P. L\'{o}pez-Sancho, and M.~C. Mu\~{n}oz, Phys.~Rev.~B {\bf 79},
  235423   (2009).

\bibitem{Jeong2009}
J.-S. Jeong and H.-W. Lee, Phys.~Rev.~B {\bf 80},  075409  (2009).

\bibitem{Izumida2009}
W. Izumida, K. Sato, and R. Saito, J. Phys. Soc. Jpn. {\bf 78},  074707
  (2009).

\bibitem{Rudner2010}
M. Rudner and E.~I. Rashba, arxiv.org/abs/1001.4306.

\bibitem{Bulaev2008}
D.~V. Bulaev, B. Trauzettel, and D. Loss, Phys.~Rev.~B {\bf 77},  235301  (2008).

\bibitem{Ando2005}
T. Ando, J. Phys. Soc. Jpn. {\bf 74},  777  (2005).

\bibitem{Golovach2006}
V.~N. Golovach, M. Borhani, and D. Loss, Phys. Rev. B {\bf 74},  165319
  (2006).

\bibitem{Laird2007}
E. Laird {\it et~al.}, Phys.~Rev.~Lett. {\bf 99},  246601  (2007).

\bibitem{Flindt2006}
C. Flindt, A.~S. S{\o}rensen, and K. Flensberg, Phys.~Rev.~Lett. {\bf 97},
  240501  (2006).

\bibitem{Trif2007}
M. Trif, V.~N. Golovach, and D. Loss, Phys.~Rev.~B {\bf 75},  085307  (2007).

\bibitem{Hu2007}
Y. Hu {\it et~al.}, Nature Nanotechnology {\bf 2},  622  (2007).

\bibitem{Biercuk2006}
M.~J. Biercuk {\it et~al.}, Physical Review B {\bf 73},  201402(R) (2006).

\bibitem{Barenco1995}
A. Barenco {\it et~al.}, Phys. Rev. A {\bf 52},  3457  (1995).

\bibitem{Belov2005}
V.~V. Belov, S.~Y. Dobrokhotov, V.~P. Maslov, and T.~Y. Tudorovskii,
  Physics-Uspekhi {\bf 48},  962  (2005).

\bibitem{Geblinger2008}
N. Geblinger, A. Ismach, and E. Joselevich, Nature Nanotechnology {\bf 3},  195
   (2008).

\bibitem{Postma2001}
H.~W.~C. Postma {\it et~al.}, Science {\bf 293},  76  (2001).

\bibitem{Bozovic2001}
D. Bozovic {\it et~al.}, App.~Phys.~Lett. {\bf 78},  3693  (2001).

\bibitem{Biercuk2004}
M.~J. Biercuk, N. Mason, J.~M. Chow, and C.~M. Marcus, Nano Letters {\bf 4},
  2499  (2004).

\end{thebibliography}
%\bibliographystyle{prsty}

\end{document}